\documentclass[preprint,12pt,authoryear]{elsarticle}

\usepackage{amssymb}

\usepackage{ulem}
\usepackage{amsmath}

\usepackage{booktabs}

\usepackage[acronym]{glossaries}
\newacronym{ai}{AI}{Artificial Intelligence}

\usepackage{url}

\usepackage{breakurl}
\usepackage[breaklinks,hidelinks]{hyperref}

\newif\ifblind
\blindfalse

\usepackage{color}
\newcommand{\blue}[1]{{#1}}

\newif\ifarXiv
\arXivtrue

\ifarXiv
    \journal{arXiv}
\else
    \journal{International Journal of Forecasting}
\fi
\begin{document}

\begin{frontmatter}



\title{The Hybrid Renewable Energy Forecasting and Trading Competition 2024}


\ifblind
\else

\author[inst1]{Jethro Browell\texorpdfstring{\corref{cor1}}{*}}
\ead{jethro.browell@glasgow.ac.uk}
\affiliation[inst1]{organization={School of Mathematics and Statistics, University of Glasgow},
            city={Glasgow},
            country={UK}}

\cortext[cor1]{Corresponding author}

\author[inst2]{Dennis van der Meer}
\affiliation[inst2]{organization={\O{}rsted A/S},
            city={Copenhagen},
            country={Denmark}}

\author[inst3]{Henrik K\"alvegren}
\author[inst3]{Sebastian Haglund}
\affiliation[inst3]{organization={Rebase Energy},
            city={Stockholm},
            country={Sweden}}

\author[inst2]{Edoardo Simioni}

\author[inst4]{Ricardo J. Bessa}
\affiliation[inst4]{organization={Centre for Power and Energy, INESC TEC},
            city={Porto},
            country={Portugal}}


\author[inst5]{Yi Wang}
\affiliation[inst5]{organization={Department of Electrical and Electronic Engineering, University of Hong Kong},
            city={Hong Kong},
            country={China}}
\fi
   
\begin{abstract}

The Hybrid Energy Forecasting and Trading Competition  challenged participants to forecast and trade the electricity generation from a 3.6GW portfolio of wind and solar farms in Great Britain for three months in 2024. The competition mimicked operational practice with participants required to submit genuine forecasts and market bids for the day-ahead on a daily basis. Prizes were awarded for forecasting performance measured by Pinball Score, trading performance measured by total revenue, and combined performance based on rank in the other two tracks. Here we present an analysis of the participants' performance and the learnings from the competition. \blue{The forecasting track reaffirms the competitiveness of popular gradient boosted tree algorithms for day-ahead wind and solar power forecasting, though other methods also yielded strong results, with performance in all cases highly dependent on implementation.} The trading track offers insight into the relationship between forecast skill and value, with trading strategy and underlying forecasts influencing performance. All competition data, including power production, weather forecasts,  electricity market data, and participants' submissions are shared for further analysis and benchmarking.

\end{abstract}

\ifarXiv

    
\fi


\end{frontmatter}


\section{Introduction}
\label{sec:intro}

Forecasting production from wind and solar power plants, and making effective decisions under forecast uncertainty, are essential capabilities in low-carbon energy systems.
Power system operators and energy traders typically rely on a combination of in-house forecasting systems and external forecasting services. Decisions informed by forecast information are generally taken by humans but are increasingly supported or automated by software.
These topics are the subject of much academic research and commercial innovation, and while both continuously report performance improvement, it is difficult to know how different approaches compare in practice. Different datasets, evaluation criteria, and the possibility of accidental (or deliberate) data leakage and misreporting all make comparisons challenging \citep{2023IEASolutions}. This problem is common across many domains and has motivated forecasting competitions to \blue{establish best practices by providing a common task and evaluation criteria, with competitions hosted by third parties to ensure accuracy and fairness} \citep{Hyndman2020AForecastingcompetitions, Hong2020EnergyOutlook}.

In the energy domain, the Global Energy Forecasting competitions of 2012, 2014 and 2017 have been particularly influential in establishing methodologies for wind, solar, price and load forecasting and for stimulating interest in probabilistic forecasting \citep{Hong2014Global2012,Hong2016ProbabilisticBeyond,Hong2019GlobalForecasting}.
Other energy forecasting competitions include those run by the European Energy Market conference, including EEM 2020, focused on national wind power production \citep{Bellinguer2020ProbabilisticApproach,Browell2020QuantileCompetition}, which has received relatively little attention in academic literature;
the impact of COVID-19 lockdowns on electricity demand motivated \citep{Farrokhabadi2022Day-AheadParadigm}, promoting the need for methods that can adapt to abrupt changes in underlying behaviours; \blue{the BigDEAL Challenge 2022 focused on short-term peak load forecasting, emphasizing the timing and shape of daily peaks, and introduced novel error metrics to benchmark models under realistic operational conditions~\citep{Shukla2024};} and two competitions on smart meter forecasting in 2020 and 2021 \citep{Pekaslan2023TheBeyond} were motivated by challenges related to billing in electricity retail. These more recent examples targeted specific challenges in energy forecasting and go beyond the mature practice of day-ahead forecasting for individual wind or solar plants, or forecasting total national/system load \citep{Hong2020EnergyOutlook}.




The Hybrid Renewable Energy Forecasting and Trading Competition (HEFTcom) was motivated by the learning and community benefits that previous competitions produced, as well as the potential for new competition to contribute to open research questions.
\blue{Specifically, the design of HEFTcom was guided by the following key objectives:
\begin{itemize}
    \item Encourage novel forecasting models tailored to combined wind and solar energy portfolios in an operational setting.
    \item Evaluate recent forecasting advances, including deep learning, versus standard methods in an operational setting for renewable energy trading.
    \item Study how forecasting accuracy impacts its value in a decision-making problem under uncertainty.
    \item Assess the complexity of the energy trading decision chain, focusing on the number of forecasting and bidding models involved in generating the optimal bid.
    \item Produce an open dataset for benchmarking.
\end{itemize}} 
\blue{First, we aim to stimulate the development of novel forecasting methods for hybrid portfolios of wind and solar power, and establish best practices for this task. Hybrid generation forecasting represents a novel aspect compared to past competitions, such as GEFcom, which focused on single technologies. As demonstrated in~\citep{Couto2023}, it poses unique challenges, such as identifying the most relevant weather parameters. Furthermore, previous competitions, notably GEFcom2014, popularised tree-based methods for wind and solar power forecasting but also highlighted the importance of forecaster expertise, data quality, preprocessing strategies, and validation techniques employed by each team~\citep{Hong2016ProbabilisticBeyond}.}

\blue{While tree-based methods have consistently performed well in structured tabular data, including in the M5 forecasting competition~\citep{MAKRIDAKIS2022M5,JANUSCHOWSKI2022trees}, recent advances in deep learning and other machine learning algorithms, especially those tailored for time series forecasting, require ongoing empirical comparison. However, the aim is not to claim superiority of one class of models over another, but to encourage researchers to systematically benchmark emerging techniques against established ones. This helps uncover potential innovations while accounting for the significant influence of forecaster expertise, data preparation, and evaluation methodology on performance.}


Secondly, the use of forecasts in decision-making and the connection between forecast performance and value are poorly understood and warrant attention. A forecast (or forecast `improvement') only has value if it leads to better decisions~\citep{2023IEASolutions,Pinson2007TradingPower}. \blue{HEFTcom aimed to advance this discussion by integrating both predictive modelling and downstream application into its structure.}
Energy trading was a natural choice for the decision-making problem as it has an inherent scoring mechanism, market revenue, and provides a link to other energy forecasting problems including price and volume forecasting. \blue{Notably, the M6 forecasting competition in financial forecasting introduced a novel evaluation approach by assessing both probabilistic forecast accuracy (e.g., Ranked Probability Score) and the effectiveness of forecasts in portfolio optimization using metrics like the Information Ratio~\citep{Makridakis2024M6,Makridakis2024b}. Inspired by such initiatives, HEFTcom encourages participants to consider not only how well forecasts perform statistically but also how they influence the decision-making outcomes.}

A further objective was to assess the complexity of \blue{the decision-making model chain, particularly the number and type of models combined to produce the submitted bid. This includes cases where multiple power and price forecasting models are integrated with decision models (e.g., stochastic optimization, heuristics rules), as well as more prescriptive approaches where a single model directly prescribes the optimal bid~\citep{Carriere2019}. Through this, HEFTcom seeks to understand on how forecasters bridge the gap between predictive analytics and actionable decisions, a topic with growing relevance in energy systems research.}

The final design consideration was the practical applicability of solutions. In practice, forecasting and decision-making in day-ahead electricity markets must be reliable and comply with fixed schedules. \blue{In contrast to past competitions,} HEFTcom was therefore set-up as a live \blue{(with daily submissions of forecasts and market bids)}, operational competition with participants forecasting future wind and solar production, and shadowing electricity market outcomes using real data from Great Britain. This set-up has the additional benefit of removing the possibility of cheating, as no restrictions were placed on the use of data beyond that provided by the competition.

HEFTcom attracted participants from around the world, including professionals working in the energy industry, students, and enthusiasts. The remainder of this paper describes and analyses the competition and its results and reflects on learnings for researchers, practitioners, and the organisers of future forecasting competitions.

\section{HEFTcom}
\label{sec:heftcom}

\subsection{Organisation}

HEFTcom was organised by the IEEE Power \& Energy Society Working  Group on Energy Forecasting and Analytics, sponsored by \O{}rsted and rebase.energy, and hosted on the IEEE DataPort \citep{Browell2024HybridCompetition}. The organising committee was 
\ifblind
[omitted for blind peer review].
\else
Jethro Browell (Chair, University of Glasgow), Sebastian Haglund (rebase.energy), Henrik K\"alvegren (rebase.energy), Edoardo Simioni (\O{}rsted), Dennis van der Meer (\O{}rsted), Ricardo Bessa (INESC TEC), and Yi Wang (University of Hong Kong).
\fi

Planning began in early 2023 with formation of the organising committee, design of the competition tasks, and construction of technical infrastructure. Competition registration, documentation and static data and a rolling leaderboard was hosted on the IEEE DataPort \citep{Browell2024HybridCompetition}, with APIs hosted by rebase.energy for data updates and submission of entries. The competition was launched  on 1 November 2023. From this date, participants were able to register and begin developing and testing their solutions.

HEFTcom was a genuine forecasting task requiring daily submissions of forecasts and market bids for the day-ahead. It was based on a hybrid generation portfolio in Great Britain comprising the Hornsea 1 wind farm and the combined solar capacity of East England, totalling approximately 3.6GW.
The main competition period was originally planned to run from the beginning of February 2024 for three months, but following a technical fault on the export cable from Hornsea 1 wind farm the competition start was delayed to give participants time to adapt to the new situation. Key competition dates were:
\begin{itemize}
    \item 1 November 2023: Competition open for registration, static data available
    \item 14 November 2023: Competition APIs and rolling weekly leaderboard open for testing
    \item 19 February 2024: First submission of the competition period (forecasts and bids for 20 February 2024)
    \item 18 May 2024: Last submission of the competition period (forecasts and bids for 19 May 2024)
    \item  24 May 2024: Deadline for participants to submit reports summarising their solutions
    \item 31 May 2024: Announcement of final leaderboard and prizes
\end{itemize}

HEFTcom had three tracks with associated prizes for the top three performing teams and best performing student team, shown in Table \ref{tab:prizes}. Live scoreboards were maintained on the competition website to provide continuous feedback to participants on their performance and were updated as data became available, which was typically with a lag of seven days. The final scoreboard was verified by the organising committee and published on 31 May 2024.

\begin{table}
    \centering
    \begin{tabular}{p{2.4cm}|c|c|c} \hline
         Rank &	Trading Track &	Forecasting Track	& Combined Ranking \\ \hline
         1st &	\$3,000	& \$3,000 & \$3,000 \\
         2nd &	\$2,000 &	\$2,000 &	\$2,000 \\
        3rd & \$1,000	& \$1,000	&\$1,000 \\
1st student &	\$1,000 &	\$1,000 &	\$1,000 \\ \hline
    \end{tabular}
    \caption{HEFTcom prizes, in USD. A team’s score in the ``Combined Ranking'' category is the sum of ranks from the trading and forecasting tracks with ties broken based on ranking in the forecasting track. Any student team finishing in the top three received the main prize and the prize for the placed student team.}
    \label{tab:prizes}
\end{table}

\subsection{Competition Data}

\begin{figure}[!t]
    \centering
    \includegraphics[width=1.0\linewidth]{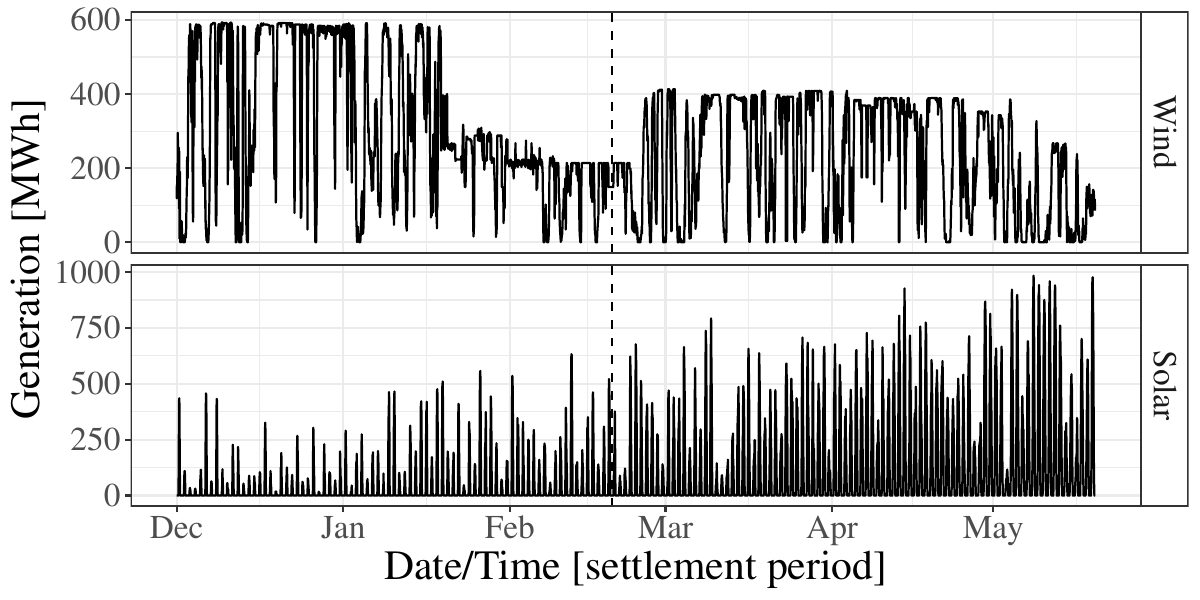}
    \caption{Wind and solar power production from 1 December 2023 to the end of the competition period on 19 May 2024. The dashed line indicates the start of the competition period. The output from Hornsea 1 was partially restricted from 19 January onwards due to an export cable fault.}
    \label{fig:wind_solar_production}
\end{figure}

HEFTcom was based on renewable generation participating in the wholesale electricity market in Great Britain (GB). HEFTcom reflected key features of this market: a day-ahead auction, half-hour settlement periods and single price imbalance settlement. Elexon is responsible for settlement in GB, which includes making relevant data publicly available. HEFTcom provided a historic dataset and simplified API for retrieving wind generation and imbalance prices from Elexon. Production data from solar is not available from Elexon as individual units are all below the size threshold that would require this, therefore we use the aggregate solar production in East England estimated by Sheffield Solar\footnote{https://www.solar.sheffield.ac.uk (Accessed 29/11/2024)}.
Wind and solar production from December 2023 to the end of the competition period is shown in Figure \ref{fig:wind_solar_production}.

\begin{figure}[!t]
    \centering
    \includegraphics[width=1.0\linewidth]{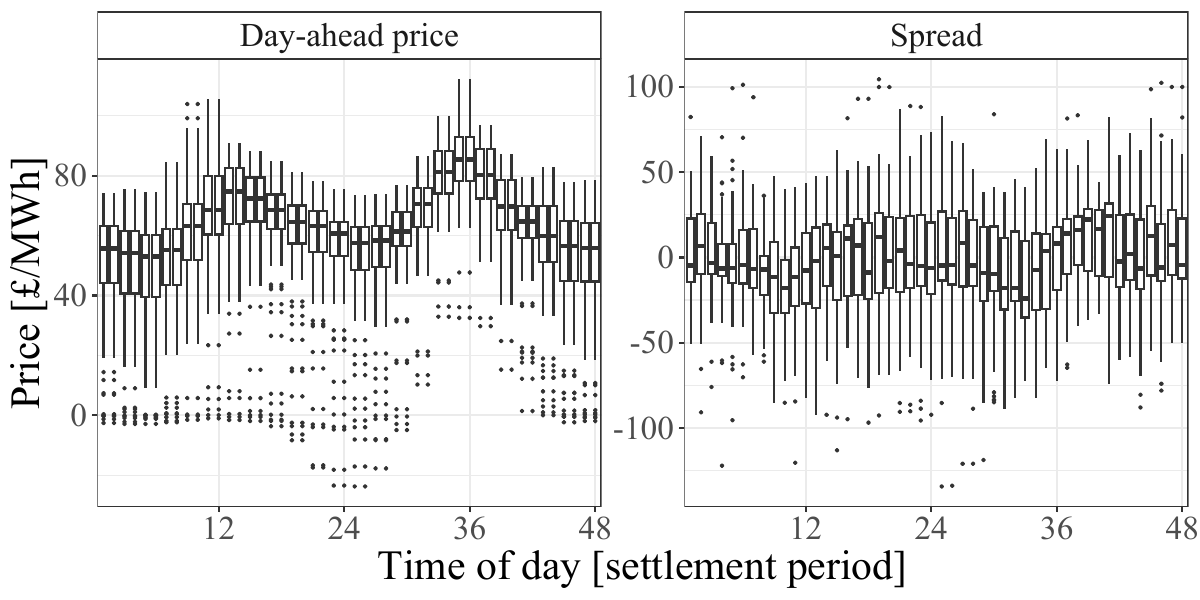}
    \caption{Box-plots of the day-ahead price $\pi_D$ and the price spread between imbalance price $\pi_S$ and $\pi_D$ during the competition period, grouped by settlement period. \blue{Boxes span the first to third quartile, whiskers extend to the largest value no further than 1.5 times the inter-quartile range, data outwith whiskers are plotted individually.} The imbalance price is typically less than the day-ahead price when the system has surplus power, and vice versa.}
    \label{fig:price_spread}
\end{figure}

There are multiple marketplaces for trading electricity in GB. To limit complexity, HEFTcom considered a day-ahead auction and imbalance settlement only. The GB single imbalance price was used directly, while the clearing price of the day-ahead auction was taken to be the `Intermittent Market Reference Price' published by the Low Carbon Contracts Company, which is the volume-weighted average price from GB's two day-ahead auctions. Box plots of day-ahead price and price spread (difference between imbalance and day-ahead price) during the competition period, grouped by settlement period, are shown in Figure \ref{fig:price_spread}.

Three years of historic and operational weather forecasts from two weather models, DWD's ICON-EU and NCEP's GFS, were made available to participants by rebase.energy. Both are hourly resolution with four updates per day. Gridded weather forecast data surrounding Hornsea 1 wind farm and East England was supplied, as well as specific points corresponding to population centres in GB (relevant for price forecasting). The use of gridded weather data has become standard practice in energy forecasting but adds significant complexity and has rarely been a feature of competitions.

A static copy of this data, along with documentation and participants' submissions, is archived at \citep{Browell2024HybridData} for benchmarking, further analysis of the competition, and reproduction of the analysis presented in this article, including the full results presented in Table~\ref{tab:full_results} in the Appendix. The HEFTcom24 GitHub repository provided a Python notebook quick-start guide to make it as easy as possible for teams to familiarise themselves with data and API formats and the competition tasks~\citep{Browell2024V1.0.0Version}.

\subsection{Participants and rules}

\begin{figure}
    \centering
    \includegraphics[width=\linewidth]{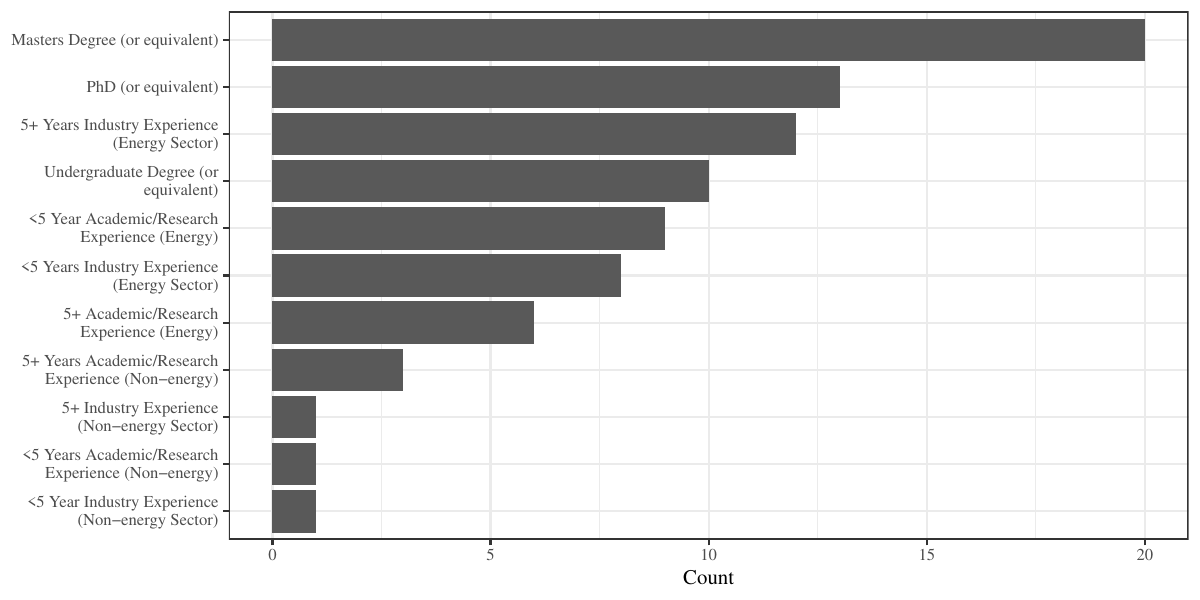}
    \caption{Skills and experience of HEFTcom participants.}
    \label{fig:team_experience}
\end{figure}

Over 170 teams registered for HEFTcom of which 66 participated by submitting at least once during the competition period.
Around two-thirds of the teams dropped out of the competition, typically after a poor performance during the early stages; ultimately, 24 teams completed the competition, including five student teams. \blue{Based on reports submitted by 37 teams, teams typically contained 1 to 4 members and most were based in Europe (29), though teams based in Asia (5), Africa (1) and North America (2) participated. Team members generally had masters or PhD degrees and industry or research experience in the energy sector, see Figure~\ref{fig:team_experience}. HEFTcom failed to attract participants from outside the energy sector.}
An online forum hosted on Slack was established to provide a line of communication between participants and organisers, which enabled fast and transparent discussion.

There was no limit on the size of teams, though all members of student teams were required to be registered students to claim a student prize. No restrictions were placed on the use of data beyond that provided by the competition. To retain a position in the final leaderboard and to qualify for a prize, participants were required to submit a report providing a high-level description of their methodology and any additional data used.
Teams were allowed to miss up to five submissions during the competition period with missed entries filled by a benchmark method. The benchmark method was provided to all teams as part of the HEFTcom24 GitHub repository \citep{Browell2024V1.0.0Version} and was listed on the leaderboard as team `Benchmark'. The organisers also contributed team `quantopia' to serve as a more competitive reference, which featured a sophisticated forecasting solution and strategic bidding algorithm\blue{, the details of which are withheld for commercial reasons}.

\section{Forecasting Track}
\label{sec:forecasting}

The forecasting track required participants to produce probabilistic forecasts of the power production from a hybrid power plant comprising the 1.2GW Hornsea 1 wind farm and the combined solar capacity of East England of approximately 2.4GW. Forecasts $\hat{q}_\alpha$ in the form of quantiles from $\alpha=10\%,~20\%,...,90\%$ for each half-hour period of the day-ahead had to be submitted by 09:20 UTC each day.

\begin{figure}[!t]
    \centering
    \includegraphics[width=\linewidth]{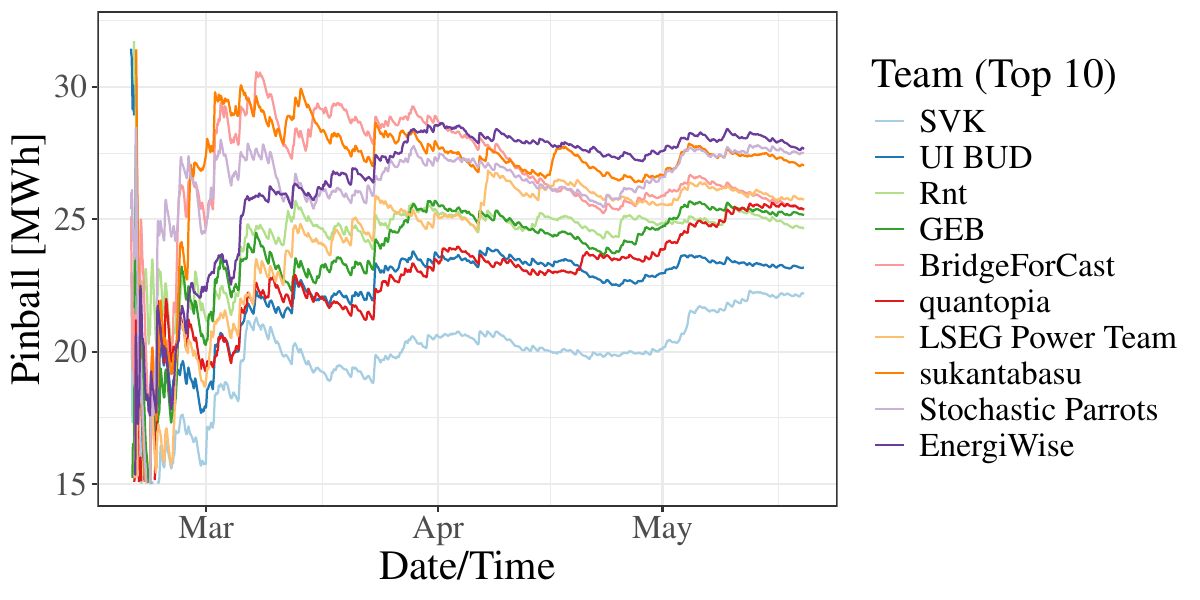}
    \caption{Expanding average of the Pinball Score for the top 10 teams in the forecasting track during the competition period.}
    \label{fig:pinball_top10}
\end{figure}

Forecasts were evaluated using the Pinball Score, \blue{the same metric used by GEFCom2014~\citep{Hong2016ProbabilisticBeyond}, defined as}
\begin{equation}
    L(y,\hat{q}_\alpha) = \begin{cases}
        ( y - \hat{q}_\alpha ) \alpha & \text{if}~y \ge \hat{q}_\alpha\\
        ( \hat{q}_\alpha - y )( 1- \alpha ) & \text{if}~y < \hat{q}_\alpha \\
    \end{cases}
\end{equation}
\blue{where $y$ is the observed value, $\hat{q}_\alpha$ is the forecasted $\alpha$-quantile, and the score is averaged over all quantiles (from 10\% to 90\%) and time steps.} The evolution of the Pinball Score for the top 10 teams in the forecasting track is shown in Figure~\ref{fig:pinball_top10}. Some common patterns are visible reflecting the variation in the predictability of wind and solar production. Specific effects are also visible, such as the seven-hour period on 23 March when Hornsea 1 wind farm did not generate due to exposure to negative wholesale prices, which none of the participants predicted. Similar events occurred on 6, 7 and 13 April; overall, 37 hours were affected by negative pricing during the competition.

Team SVK established an early lead that was maintained until the end of the competition despite a period of relatively poor performance in early May (caused by human error~\citep[ISF presentation]{Browell2024HybridData}). Other rankings changed frequently, but rarely was a team more than two positions away from their final rank after the first month. BridgeForCast is a notable exception, who recovered from 13th position after one month to finish 5th, posting the best performance of all teams in the final two months of the competition.

While this setting is similar to past competitions and many academic studies, the objective of forecasting the total production from a mixed portfolio presents a new challenge, as did the operational nature of the competition. Two practical aspects in particular impacted the competition, the export cable fault at Hornsea 1 wind farm and technical issues causing NWP data to be delayed.

The cable fault occurred on the morning of 19 January, and its impact is clearly visible in Figure~\ref{fig:wind_solar_production}. The fault was reported publicly via REMIT, Regulation (EU) No 1227/2011 on wholesale energy market integrity and transparency, which requires market participants to share plant availability information. However, neither the competition organisers nor participants were monitoring this data feed, and it wasn't until after the competition had initially started on 1 February that the issue was identified. While the objective of running the competition live was to encourage participants to be robust and respond to unexpected events, in practice, forecasters would be aware of REMIT, and it was an oversight of the organisers not to anticipate this possibility; therefore, the competition was restarted on 20 February. Ultimately, 64 REMIT messages related to the export limit for Hornsea 1 were published between 19 January and the end of the competition on 20 May.

\begin{figure}
    \centering
    \includegraphics[width=\linewidth]{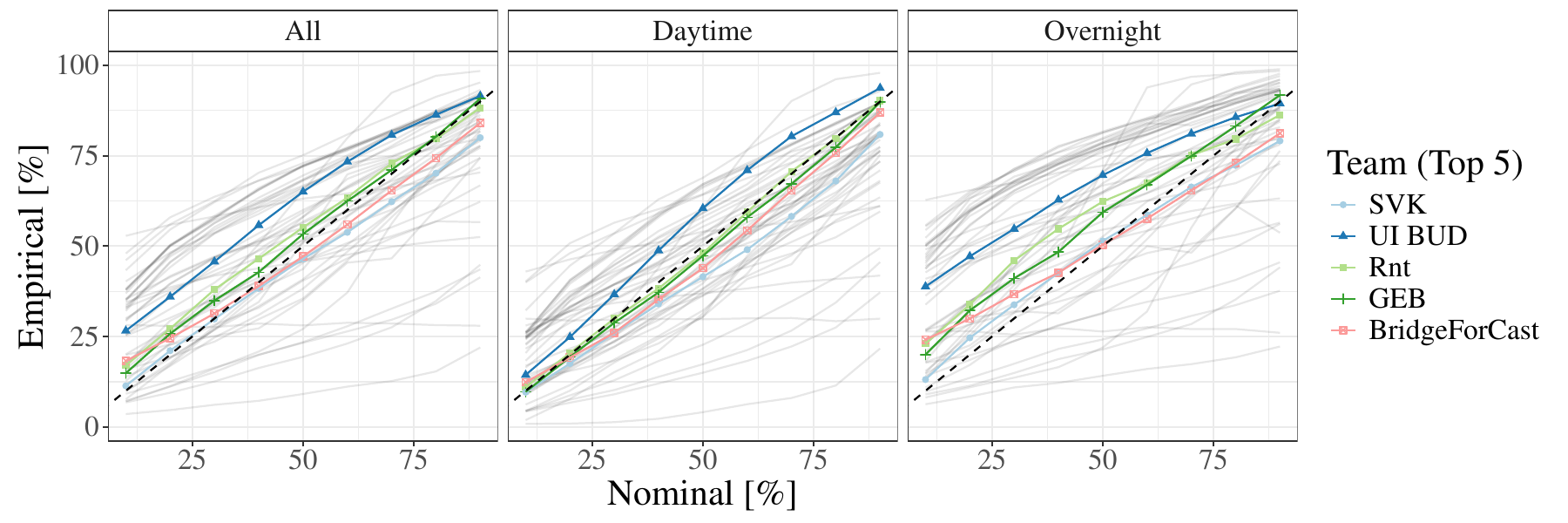}
    \caption{Reliability diagrams for HEFTcom submissions from participants who submitted on at least 50\% of days. Daytime, defined here as 8am--8pm UTC, and overnight periods are separated to compare periods of wind-only production (hours of darkness) with wind and solar. The top-5 teams in the forecasting track are highlighted.}
    \label{fig:reliability}
\end{figure}

Probabilistic forecasts should be calibrated, which is to say that the frequency of events should match the probability with which they are predicted. Calibration can be assessed via reliability diagrams, which compare empirical frequency with nominal/predicted probability. Reliability diagrams for HEFTcom submissions are shown in Figure~\ref{fig:reliability}. While most of the top-5 teams produced calibrated forecasts, UI BUD remarkably achieved a competitive average Pinball Score with substantial bias;
\blue{however, it should be noted that calibration and sharpness may be traded-off if the objective is minimisation of Pinball Score \citep{Candille2005EvaluationCRPSdecomp}, though in general calibrated forecasts are preferred \citep{Gneiting2007ProbabilisticSharpness}.}

\blue{Daytime, defined here as 8am--8pm, and overnight periods are separated to isolate period of wind-only production (hours of darkness) and compare to periods of wind and solar production. The daytime Pinball Score of most teams is approximately double that of overnight reflecting how pinball scales with the level of total generation. UI BUD in fact have the lowest daytime Pinball Score of participants, narrowly beating SVK in second, but SVK has a substantially lower Pinball Score than UI BUD overnight. Pinball Scores for the top-10 teams in the forecasting track separated by day/night are listed in Table~\ref{tab:pinball_time_of_day}}.
Also notable is the cluster of participants who, similar to UI BUD, consistently over-forecast, especially during the night, suggesting that wind power is being over-predicted. This is possibly related to the reduced export capacity of Hornsea 1 and highlights the challenge the participants faced with the training data, not including similar periods of constrained production.
\begin{table}[t]
    \centering
    \caption{\blue{Pinball scores for the top-10 teams in the forecasting track averaged over all time periods, daytime, defined as 8am--8pm, and overnight. All units are MWh.}}
    \label{tab:pinball_time_of_day}
    
    \begin{tabular}{lccc}
          \hline
        Team & All & Daytime & Overnight \\ 
          \hline
        SVK & 22.18 & 30.88 & 13.48 \\ 
          UI BUD & 23.18 & 30.60 & 15.74 \\ 
          Rnt & 24.64 & 31.93 & 17.35 \\ 
          GEB & 25.16 & 33.32 & 16.99 \\ 
          BridgeForCast & 25.34 & 33.30 & 17.38 \\ 
          quantopia & 25.38 & 35.88 & 14.86 \\ 
          LSEG Power Team & 25.74 & 34.93 & 16.55 \\ 
          sukantabasu & 27.04 & 34.89 & 19.17 \\ 
          Stochastic Parrots & 27.50 & 36.68 & 18.32 \\ 
          EnergiWise & 27.65 & 33.89 & 21.41 \\ 
           \hline
    \end{tabular}
\end{table}

\begin{figure}
    \centering
    \includegraphics[width=\linewidth]{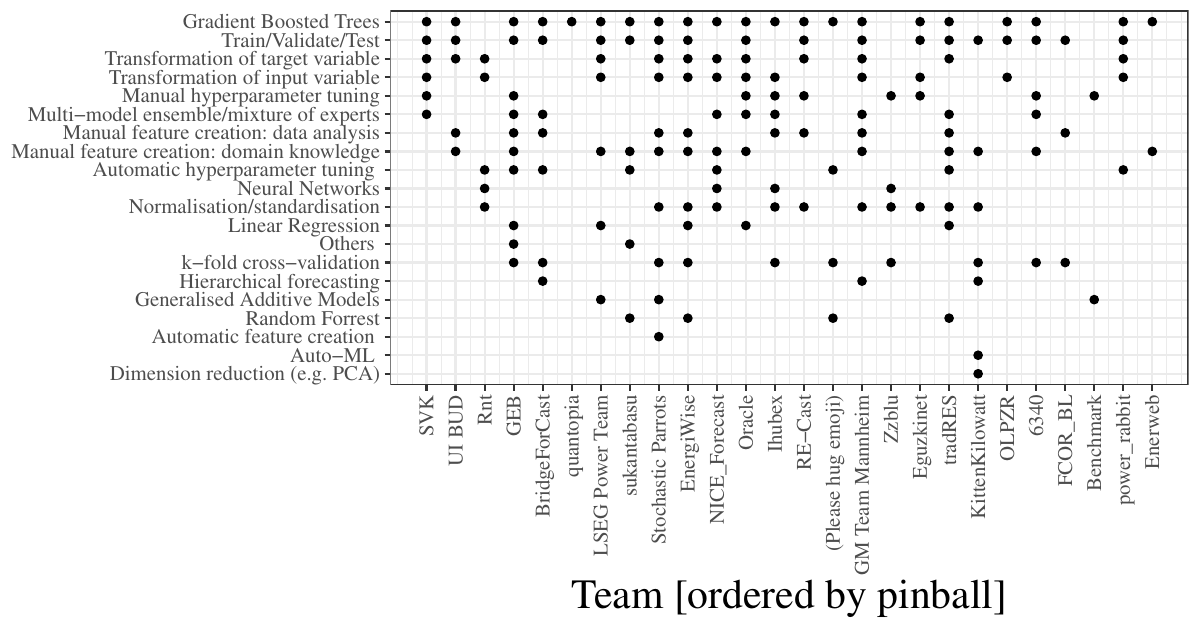}
    \caption{Methods used by teams in the forecasting track. Some details for \textit{quantopia} were withheld \blue{for commercial reasons}.}
    \label{fig:forecast_methods}
\end{figure}

The methods used in the forecasting track, as reported by participants, are summarised in Figure~\ref{fig:forecast_methods}. Common features across top-performing teams are the use of Gradient Boosting Trees, the combination of multiple models, feature selection and hyper-parameter tuning. However, the fact that the majority of teams used most, if not all, of these methods highlights the importance of their implementation. For example, top-performing teams selected features based on training/validation experiments or feature importance, whereas lower-ranked teams selected features based on exploratory data analysis. \blue{75\% of teams, including nine of the top 10, forecast wind and solar separately and then combined these forecasts using either an additional model or quantile aggregation scheme.}

Several teams, including SVK, Rnt and BridgeForCast, used additional weather forecast data beyond what was provided by the competition. SVK reported an 8\% improvement in Pinball Score after combining forecasts from the MET Norway's MetCoOp Ensemble Prediction System\footnote{https://thredds.met.no/thredds/metno.html (Accessed 21/11/2024)} with the GFS and DWD forecasts provided by the competition, based on analysis using 2023 as a validation set. However, two teams in the top-5\blue{, UI BUD and GEB,} did not use additional weather forecast data.

Rnt is the most distinctive of the top-10 as their approach did not involve tree-based methods, instead using embeddings from in-house \blue{\gls{ai}} weather models as input to downstream neural networks that predicted solar and wind generation. The \gls{ai} weather models were based on those described in \citep{Andrychowicz2023DeepObservations} but extended to include solar irradiance and day-ahead lead-times. Input data included observation data from weather stations, radar and satellite imagery, and NWP analysis.

Approaches to handling the cable fault varied according to the reports submitted by participants; none reported any handling of negative pricing. Some re-scaled or clipped/capped predictions, while others re-trained models on clipped/capped training data. Despite the restarting of the competition one month after the initial cable fault, many participants struggled to adapt in the early stages of the competition. Participants who were forecasting total wind and solar production directly found it harder to adapt than those who could apply simple post-processing methods to their wind power forecast directly. The Benchmark did not account for the cable fault and performed extremely poorly as a result.

\blue{The winning approach of SVK is described in detail in \citep{Olauson2025}, and may be summarised as follows. CatBoost models (gradient boosting decision trees) were fit for each source of NWP (DWD, GFS, and MEPS) separately, and independently for wind and solar using the \texttt{MultiQuantile} loss targeting the nine required quantiles. The features used were the NWP grid points, raw, lagged and differenced, plus calendar features. The only hyper-parameter tuned was the number of boosting iterations, default values of all others were used and unimportant features were dropped after initial testing. Quantiles were clipped to the maximum capacity accounting for outages given by REMIT. Next, meta-models were used to combine the CatBoost model predictions for wind and solar separately; these comprised a linear quantile regression model for each target quantile with all 27 predicted quantiles from the three CatBoost models as covariates. Finally, quantile predictions for wind and solar were added together by quantile with an adjustment for correlation between wind and solar; however, testing revealed that this adjustment yielded only a very small benefit. If ever NWP data was missing, the quantile predictions for its corresponding CatBoost model would be missing and were filled with predictions from available models, providing a level of robustness to missing input data.}

\section{Trading Track}
\label{sec:trading}

The trading track required participants to sell the energy produced by the hybrid portfolio in the day-ahead electricity market subject to imbalance settlement. This track is based on Great Britain’s wholesale electricity market, which features a day-ahead auction and single-price imbalance settlement. GB's intraday auctions and power exchange (continuous bilateral trading) were not included in HEFTcom.

For each 30-minute settlement period, total revenue $R$ is the sum of revenue from the day-ahead market based on volume $x$ sold at day-ahead price $\pi_D$, and revenue from the imbalance market. The imbalance market settles the difference between traded energy $x$ and actual production $y$. In practice, a market participant’s behaviour can influence both the day-ahead and imbalance price, but here we assume that participants are price-takers in the day-ahead market, and the price-maker effect in the imbalance market is modelled.

A market participant’s own imbalance volume, the difference between their actual generation $y$ and traded volume $x$, will influence the system’s net imbalance volume and, therefore, the imbalance price.
We model this effect for the purpose of the competition by calculating an imbalance price for each participant based on the actual Single System Price $\pi_S$ and the participant's imbalance volume.
A participant's effective imbalance price is given by $\pi_S - 0.07(y-x)$, where $-0.07$ is the regression coefficient between the net imbalance volume and imbalance price calculated from recent historical data and, therefore, represents the average impact of changes in imbalance volume on the Single System Price. This represents a simplification as, in practice, the relationship between net imbalance volume and imbalance price is non-linear and uncertain at the day-ahead stage.

For each half-hour settlement period, revenue is calculated as
\begin{equation}
    R = x \pi_D + ( y - x ) \left( \pi_S - 0.07 ( y - x ) \right) \quad ,
    \label{eq:revenue}
\end{equation}
where $x \times \pi_D$ is revenue from the day-ahead auction, $y - x$ is the participant's imbalance volume, and $\pi_S - 0.07 \times ( y - x )$ is the participant's imbalance price.
The quadratic nature of the revenue $R$ allows us to calculate the optimal trade $x_{\text{opt}}$ as
\begin{align}
    x_{\text{opt}} &= y - \frac{\pi_S - \pi_D}{0.14} \quad , \label{eq:optimal_bid}
\end{align}
where $\pi_S - \pi_D$ is referred to as the price spread between the imbalance and day-ahead markets. All three quantities on the right-hand side of \eqref{eq:optimal_bid} are unknown at the time market bids are submitted. Participants are already forecasting $y$ in the forecasting track; handling the price spread represents an additional prediction challenge for teams wishing to bid strategically. \blue{Substituting \eqref{eq:optimal_bid} into \eqref{eq:revenue} yields the theoretical maximum revenue possible in the competition.}

Equation \eqref{eq:optimal_bid} implies that bidding the expected production $x = E[y] \approx \hat{q}_{50\%}$ maximises revenue in expectation only when the price spread is zero. The bid that maximizes expected revenue is greater than $\hat{q}_{50\%}$ when the spread is negative, and less than $\hat{q}_{50\%}$ when the spread is positive. However, the price-maker effect means that large imbalance volumes are penalised regardless of price spread.

\begin{figure}[!t]
    \centering
    \includegraphics[width=\linewidth]{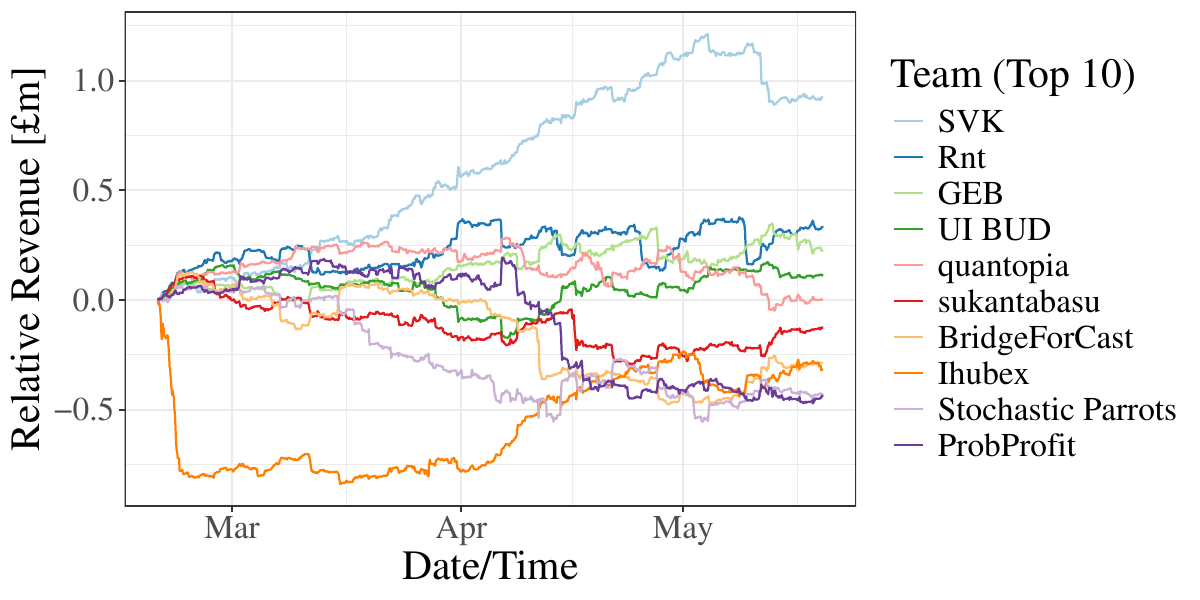}
    \caption{Rolling revenue of the top-10 teams in the trading track less the mean revenue of the top-10. SVK established a lead after around four weeks which they retained until the end of the competition by consistently out-performing other competitors in the top 10, while Ihubex had a poor start but performed well enough in the first two weeks of April to secure a top-10 finish.}
    \label{fig:revenue_top10}
\end{figure}

The evolution of participants' revenue relative the the mean of the top-10 finishers in the trading track is shown in Figure~\ref{fig:revenue_top10}. This track was much more volatile than forecasting. As in the forecasting track, SVK established a lead in March which was maintained until the end of the competition, though the performance of competitors was less consistent. Unlike in the forecasting track, where competitors' performance was highly correlated, in trading, the behaviour was much more diverse, reflecting the greater variety of methods and strategies employed by different teams.

\begin{figure}
    \centering
    \includegraphics[width=0.7\linewidth]{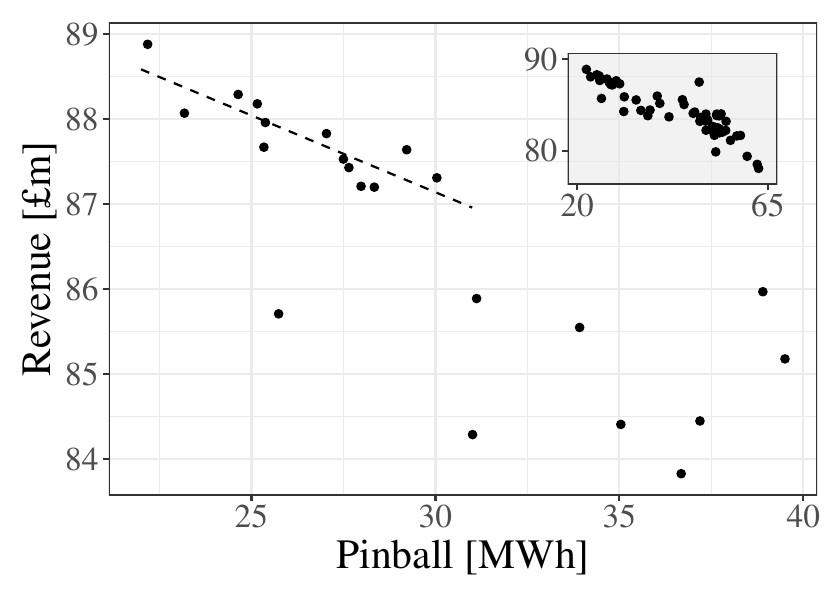}
    \caption{Scatter plot of Pinball Score vs Revenue for the  22 teams with Pinball Scores less than 40 MWh, \blue{regression line fit to teams with Pinball less than 31 MWh excluding outlier at (25.7, 85.7),}  and inset showing all teams omitting five outliers.}
    \label{fig:pinball_vs_rev}
\end{figure}

Performance between the forecasting and trading tracks was highly correlated, but rank correlation is not perfect, illustrated in Figure \ref{fig:pinball_vs_rev}. Performance in the trading track depends on both forecast skill and the effectiveness of trading strategy, which explains some of the variation. Teams with successful bidding strategies were able to exceed expectations based on their Pinball Score, while others were heavily penalised for poor trading strategy. \blue{The significance of this relationship is verified through simple linear regression of Revenue on Pinball Score, excluding outliers and teams with a Pinball Score greater than 31 MWh. The fit has gradient $-0.18$ \pounds m/MWh with 95\% confidence interval $(-0.25,-0.11)$, verifying significance at that level $p<0.001$.}

\begin{figure}[!t]
    \centering
    \includegraphics[width=\linewidth]{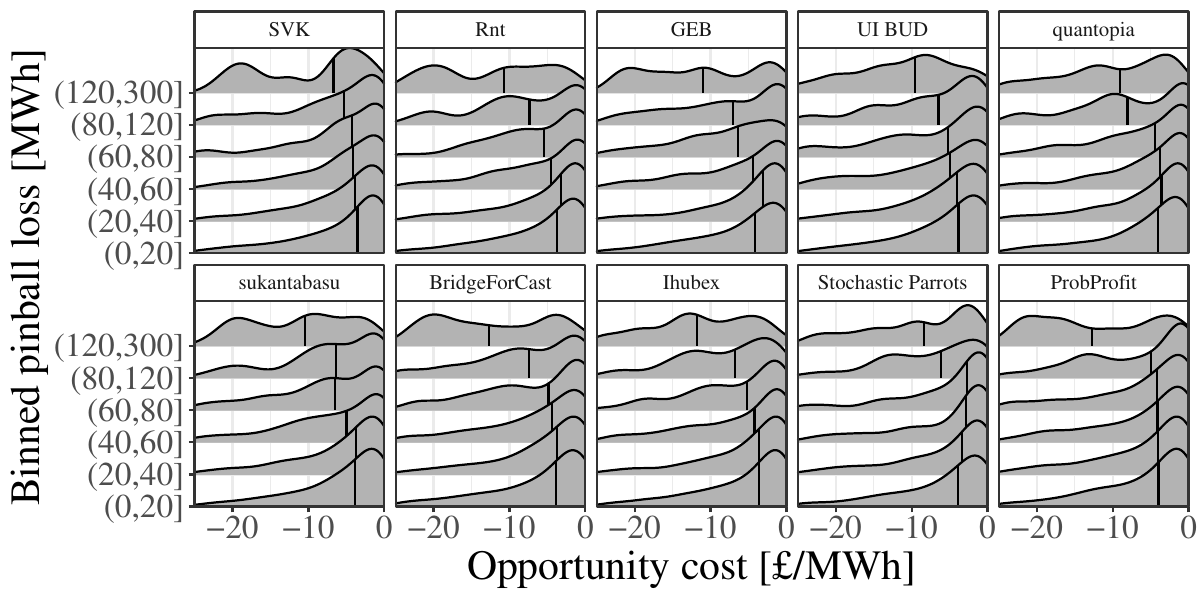}
    \caption{Opportunity cost, defined as the revenue minus maximum achievable revenue per unit volume traded, versus the Pinball Score, presented as ridgeline plots to visualize the distribution of the data\blue{. Vertical lines indicate the median}.}
    \label{fig:opportunity_cost}
\end{figure}

First, we will analyse the effectiveness of participants' trading strategies by considering how much of the theoretical maximum revenue they were able to capture. Consider opportunity cost per MWh traded in each settlement period, defined as actual revenue minus theoretical maximum, normalised by trade volume.
Figure~\ref{fig:opportunity_cost} presents the opportunity cost binned Pinball Score for corresponding periods.
We observe that the median opportunity cost is typically around 5 \pounds/MWh, but with a long tail, even for periods with low Pinball Scores (accurate forecasts).
The top teams in the trading track are differentiated by how well they were able to capture revenue during periods that were more challenging to forecast, and the frequency of large losses. 

Revenue capture by teams SVK, Stochastic Parrots, Ihubex\blue{, quantopia} and ProbProfit appear less affected by large forecast errors, as indicated by the consistent median opportunity cost \blue{and fewer large costs associated with relatively poor forecasts that had Pinball Losses in the range 40--80 MWh}. As we will see, these teams engaged in strategic bidding using, directly or indirectly, information about the price spread. However, GEB also bid strategically but do not fit this pattern. Rnt, UI BUD, BridgeForCast and sukantabasu generally bid their $\hat{q}_{50\%}$ forecast and suffered relatively large opportunity costs during periods of poor forecasting compared with the strategic bidders. \blue{All teams experienced large opportunity costs associated with forecasts with Pinball Losses greater than 80 MWh.}

\begin{figure}[!t]
    \centering
    \includegraphics[width=\linewidth]{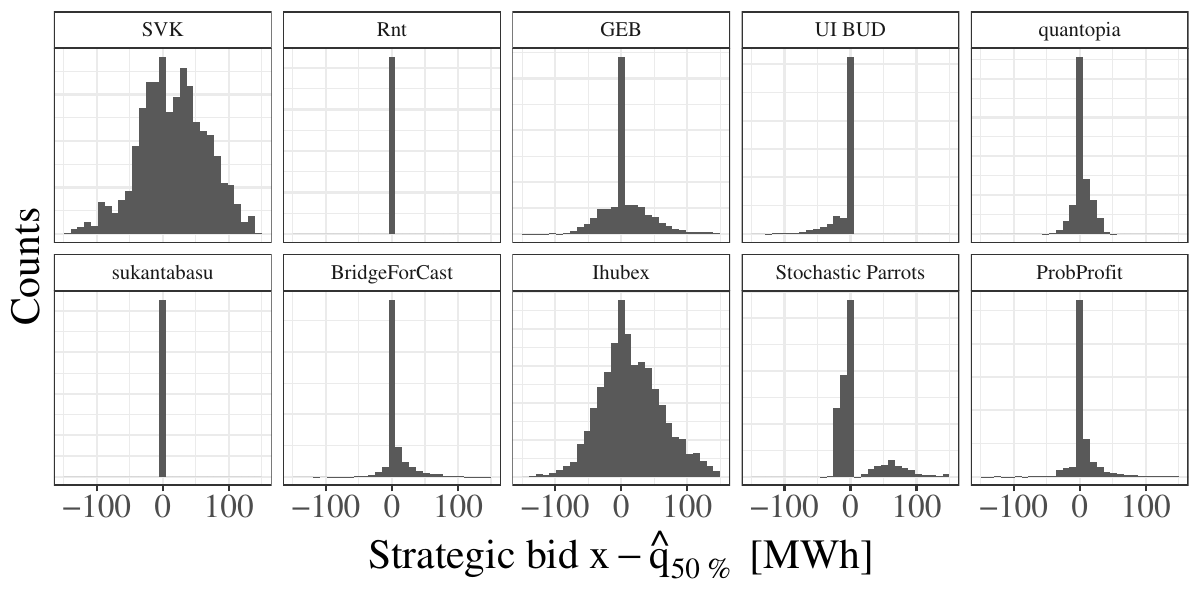}
    \caption{Histograms of strategic bid volumes ($\hat{q}_{50\%}-x$) for the top-10 teams in the trading track. Some chose to bid their $\hat{q}_{50\%}$ forecast the majority of the time, while those employing algorithmic trading strategies had more diverse bids. Teams GEB, UI BUD, and BridgeForCast began strategic bidding part way through the competition.
    }
    \label{fig:bidhist_top10}
\end{figure}

The degree of strategic bidding varied substantially between teams with some bidding $x=\hat{q}_{50\%}$ the majority of the time, and others being much more dynamic, as illustrated in Figure~\ref{fig:bidhist_top10}. Several teams changed strategy midway through the competition.
Note that the optimal bid \eqref{eq:optimal_bid} can be thought of as an imbalance in the opposite direction to the price spread equal to $y-x = -\frac{\pi_S-\pi_D}{0.14}$.
The success of bidding strategies is, therefore, highly correlated with how frequently participants' imbalance volume was in the opposite direction to the price spread, regardless of a particular strategy.
The most successful team who primarily bid their $\hat{q}_{50\%}$, Rnt, finished second in the trading track and their imbalance was opposite to the price spread 48.9\% of the time, two percentage points higher than other teams following this strategy.

Teams who bid strategically were able to increase the rate at which they bid in the correct direction relative to their $\hat{q}_{50\%}$. SVK bid in the correct direction 56.0\% of the time resulting in their imbalance being opposite to the price spread in 51.5\% of settlement periods, higher than any other team. After a poor start, Ihubex performed well with the highest revenue during April--May. They bid in the correct direction 57.5\% of the time, resulting in an imbalance opposite to the price spread in  50.3\% of settlement periods.

\begin{figure}[!t]
    \centering
    \includegraphics[width=\linewidth]{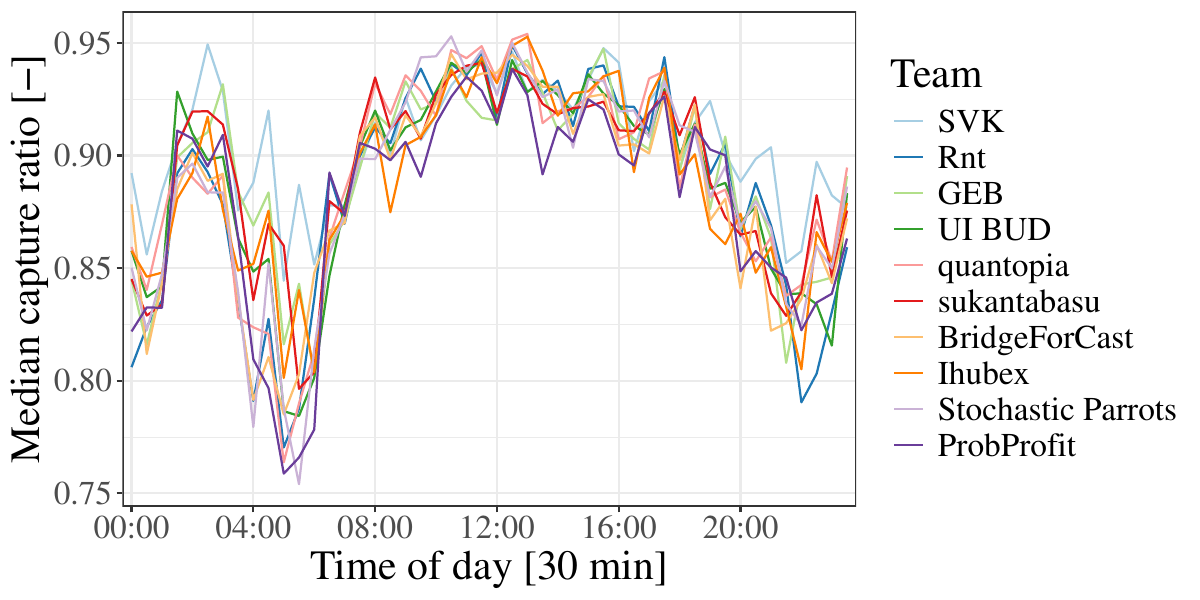}
    \caption{Median capture ratio of the top 5 teams, defined as the revenue achieved by the team divided by the maximum revenue as per \eqref{eq:optimal_bid}.}
    \label{fig:capture_ratio}
\end{figure}

We are also interested in the times of day at which teams were able to capture the most value from their assets, which is illustrated in
Figure~\ref{fig:capture_ratio}. There is a clear trough for most teams around 05:00 in the morning and the median capture ratio is similar across teams during midday and the highest during the entire day. Notably, SVK outperforms the rest of the top-5 in select hours during the early morning and late evening when the price spread is systematically negative and positive, respectively, as can be seen in Figure~\ref{fig:price_spread}.

Besides accurately forecasting energy production, \eqref{eq:optimal_bid} implies that forecasts of $\pi_D$ and $\pi_S$ (or at least $\pi_D - \pi_S$) are necessary to determine $x_\text{opt}$, and this was the approach followed by Ihubex, GEB and others~\citep{Browell2024HybridData,pu2025heftcom2024}. Ihubex forecasted prices using multiple models and applied a set of heuristics to balance risk and revenue. GEB produced probabilistic forecasts of the price spread directly and combined this with their generation forecasts to find the bid that maximised expected revenue \citep{pu2025heftcom2024}. SVK, on the other hand, transformed the decision problem into a prediction problem by creating a historical dataset of optimal bids and training a gradient boosted decision tree to predict the optimal bid directly using their generation forecasts, calendar variables, and average price statistics as features \citep{Olauson2025}.

Two different routes to success in the trading track are apparent. First, attempting to minimise imbalance volumes using a forecast that is less correlated with the price spread than other competitors. The price spread is influenced by wind power forecast errors in the GB market overall, so it is notable that Rnt, who had success via this route, had a distinctive forecasting approach based on an \gls{ai} weather model not widely used in practice. The second successful approach was to bid strategically to increase revenue from imbalance settlement. The most successful strategic bidders modulated the size of their expected imbalance to maximise revenue in the long-run, though this resulted in greater volatility in their revenue. Of the teams that engaged in strategic bidding, only Ihubex, SVK and GEB increased revenue relative to what they would have achieved by bidding their $\hat{q}_{50\%}$ by more than \pounds 500,000 overall, equivalent to 0.31 \pounds/MWh of production. The next closest teams achieved gains of only \pounds 150,000 or less.

These differences manifest in summary statistics of revenue calculated over each period of the competition, listed in Table~\ref{tab:trade_statistics}. Differences are also apparent in the risk profile of trading strategies. Most strategic bidders sold more energy than the hybrid portfolio produced in the day ahead market, taking short positions on average, with relative bid volumes of 1.01--1.04. This is a risky strategy as it is expensive to buy back a deficit if the system as a whole is in deficit causing the imbalance price to be very high, a much greater penalty than the modest return received when the deficit is bought back at an imbalance price slightly lower than the day-ahead price. This is reflected in the 5\% Value at Risk (the 5\% quantile of revenue) for teams SVK and Ihubex in particular, which is very negative compared to other top-performing teams. quantopia stand out for favouring long positions and the use of strategic bidding to reduce risk while achieving a competitive final revenue.
\begin{table}[t!]
\centering
\caption{Table with common trade statistics of the top-10 teams in the trading track. Win rate, the proportion of periods with positive revenue; Volume Weighted Average Price (VWAP) for volume bid in the day-ahead market; VWAP for actual generation; Sharpe and Sortino ratios; Value at Risk (VaR), the 5\% quantile of revenue by period; and Expected Shortfall, the mean of revenue by period below the 5\% quantile.} \label{tab:trade_statistics}
\resizebox{\textwidth}{!}{%
\begin{tabular}{lrrrrrrrr}
  \hline
Team & Win rate & Relative bid volume & Trade VWAP  & Production VWAP & Sharpe ratio & Sortino ratio   & 5\% VaR & 5\% ES \\
Unit & [\%] & [-] & [\pounds/MWh] & [\pounds/MWh] & [-] & [-] &  [\pounds] & [\pounds] \\ \hline
SVK & 92.6 & 1.01 & 54.09 & 54.51 & 1.286 & 3.122 & -591.13 & -4546.66 \\ 
  Rnt & 94.1 & 0.99 & 54.48 & 54.15 & 1.259 & 1.977 & -43.08 & -4657.02 \\ 
  GEB & 93.4 & 1.01 & 53.35 & 54.08 & 1.263 & 2.735 & -114.49 & -4346.87 \\ 
  UI BUD & 93.6 & 1.03 & 52.54 & 54.01 & 1.253 & 2.241 & -119.73 & -4532.78 \\ 
  quantopia & 93.0 & 0.96 & 55.93 & 53.95 & 1.277 & 2.987 & -25.95 & -3430.61 \\ 
  sukantabasu & 93.5 & 1.02 & 52.89 & 53.87 & 1.256 & 2.443 & -248.14 & -4916.53 \\ 
  BridgeForCast & 93.5 & 1.01 & 53.43 & 53.77 & 1.262 & 2.799 & -157.63 & -3975.42 \\ 
  Ihubex & 92.3 & 1.04 & 51.82 & 53.75 & 1.238 & 3.011 & -818.36 & -5517.93 \\ 
  Stochastic Parrots & 93.7 & 1.02 & 52.77 & 53.68 & 1.249 & 2.765 & -149.21 & -4129.88 \\ 
  ProbProfit & 94.2 & 0.95 & 56.40 & 53.68 & 1.279 & 3.098 & -94.89 & -3647.48 \\ 
   \hline
\end{tabular}
}
\end{table}

\begin{figure}
    \centering
    \includegraphics[width=0.7\linewidth]{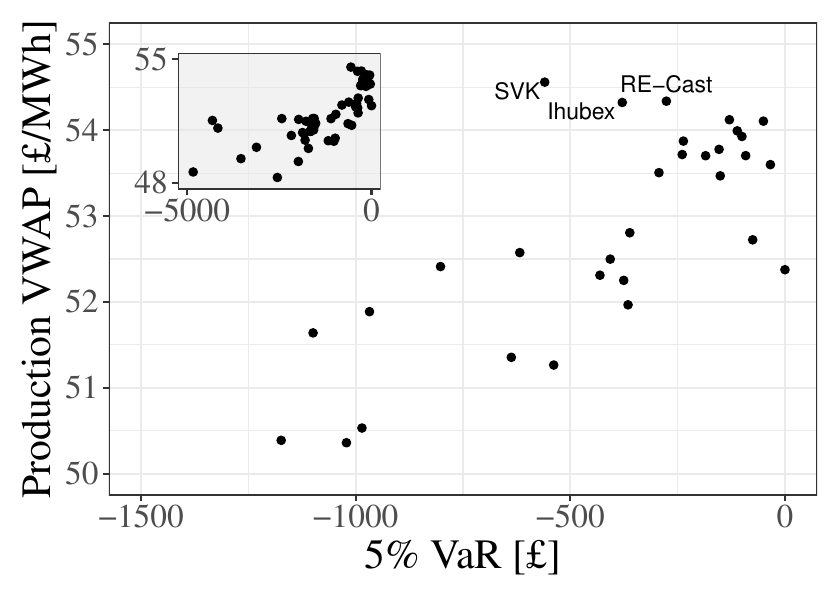}
    \caption{Risk vs average price per unit of production achieved excluding the first week of the competition where some teams had poor/unrepresentative performance. Three teams (labelled) were able to buck the trend and convert risk into reward. All teams are shown in the inset except for three outliers with very low VWAP and VaR.}
    \label{fig:risk_vs_revenue}
\end{figure}

Amongst all competitors, there is a general trend that taking greater risk resulted in lower revenue. However, SVK, Ihubex, and RE-Cast were able to increase their overall revenue alongside a modest increase in risk (in individual periods) against this trend. These teams effectively increased their expected revenue per period at the expense of increasing the variance of returns, therefore achieving greater revenue in the long run. 
This is illustrated in Figure~\ref{fig:risk_vs_revenue} and is also reflected in the Sharpe and Sortino ratios\footnote{\blue{The Sharpe ratio is a measure of risk-adjusted returns given by mean revenue divided by standard deviation of revenue. The Sortino ratio is a measure of risk-adjusted returns considering only down-side volatility given here by mean revenue divided by the standard deviation of negative revenues.}} in Table~\ref{tab:trade_statistics}. 
Such strategies rely on accurate forecasts of both market conditions and production to increase the frequency of profitable surplus/deficit positions and the returns made in those periods.

\section{Discussion}
\label{sec:discussion}

HEFTcom broadly succeeded in its aims to (re-)establish \blue{best practices} for renewable energy forecasting and to promote decision-making problems that are intertwined with forecasting. The high level of engagement from industry and academia in both organising and participating in HEFTcom highlights the relevance of the problems it addressed, and the role data science competitions can play in stimulating research and professional development \citep{Orsted2024Innovation}. However, we did not succeed in attracting new ideas from other fields, with almost all participants working or studying in the energy sector. Previous energy forecasting competitions were more successful in this regard, perhaps benefitting from being hosted on data science competition platforms, such as Kaggle and crowdANALYTIX in the case of GEFcom 2012 and 2014, respectively, whereas HEFTcom was hosted on IEEE DataPort, which is energy and engineering focused.

Establishing \blue{best practices} for renewable energy forecasting is valuable for both practitioners and the research community. Implementing forecasting systems in practice requires significant investment, which must be justified by a realistic expectation of performance and value, which HEFTcom provides. The potential of \gls{ai} weather models in the energy sector is particularly tantalising. For researchers, any new forecasting methodology should be compared to the current state-of-the-art and ideally on an open dataset to enable results to be reproduced. HEFTcom adds to the growing number of benchmark datasets for energy forecasting and is more comprehensive than its predecessors \citep{Browell2024HybridData}.

The results of the trading track offer a detailed exploration of the link between forecast skill and value in energy trading for the first time. While the necessary simplifications to the trading problem mean the results must be interpreted carefully, they nevertheless provide a signal that marginal gain in forecast skill can be of financial value. Results also show that the effectiveness of decision-making under forecast uncertainty is of similar importance to forecast skill, especially as forecast improvement alone is expected to yield diminishing return on investment. \blue{Teams that successfully engaged in strategic bidding added over \pounds500,000 to their revenue compared the naive strategy of bidding their $q_{50\%}$. Using the regression analysis above, this is equivalent to a forecast improvement of over 10\%, which is substantial. Moreover, bidding a perfect deterministic forecast would have resulted in a revenue of \pounds92.0m, only \pounds3.2m more than the winning team, while perfect decision-making would have generated \pounds105.2m. When allocating limited resources, practitioners should carefully consider the prospects of achieving gains in one or both areas relative to their current capability, and balance this against the effort required.}

Running HEFTcom as a live competition was necessary to enable the trading track to follow a real electricity market and to allow participants to incorporate additional data beyond that provided by the organisers. Being a genuine forecasting problem also provides trust in results as data leakage, accidental or deliberate, was impossible. Additionally,  participants (and organisers) were forced to consider practical issues, such as being robust to missing data, technical problems, and unexpected events, which are critical in forecasting practice but often overlooked by academic studies and competitions. \blue{HEFTcom required 90 submissions on consecutive days, which is a significant burden, though automation of API submissions was supported to lessen this. The number of missed submissions is included alongside full results in the Appendix. Of the top 15 finishing teams, nine missed no submissions, five missed one, and one missed two. Teams with high numbers of missed submissions were those that abandoned the competition part way through.} \blue{Some teams reported making minor errors during the competition, but none were as consequential as ProbProfit's erroneous submission affecting only their 40\% quantile for 2024-05-16 22:00:00, but producing a Pinball Score of the order $10^{14}$ MWh. The trading track was unaffected. Excluding this erroneous forecast, ProbProfit's average Pinball Score is a respectable 29.47 MWh.}

Feedback from participants who completed the competition was extremely positive, with many teams enjoying and learning from the experience, though this was a self-selecting sample of participants who fared well in the competition. Several commented that the time required to develop a solution was significant and, as a result, they underperformed in the early stages. The organisers debated the balance between realism, complexity, and incentive to participate and recognise that there is no perfect solution. With the increasing number of data science competitions available, competition organisers are now in competition for participants and must be mindful of the investment required to participate. HEFTcom perhaps failed to attract participants from outside of the energy space because it was an unattractive prospect to non-specialists: potential barriers to entry included the need for automation,  specialist weather data formats, and the complexity of energy systems and markets. It is worth noting that several top-performing teams were experienced in operational energy forecasting. This makes the strong performance of student teams in HEFTcom all the more impressive.

Organisers of future energy forecasting competitions will have to consider many of these issues and be guided by their aims, such as stimulating activity on a particular research problem or serving as a tool for education, publicity, or some other purpose. Increasing access to open data and cloud infrastructure has dramatically lowered the barriers to running live competitions, which have many advantages, not least enhancing the integrity and realism of the competition. Formats based on submitting either data or software may suit different settings according to need. For example, HEFTcom required daily submission for over three months, which was a substantial burden. Competition duration must be sufficient to produce meaningful results but short enough to not deter participation. Despite a two-month testing phase and restarting the competition following the Hornsea 1 cable fault, many teams struggled in the first few days of HEFTcom, and others had spells of poor performance. For this reason, formats that include multiple rounds and prizes, as in the M6 competition \citep{Makridakis2024M6}, are attractive, but necessitate longer competitions so that ranking in each round is reflective of skill rather than luck.

Future energy forecasting competitions should focus on current and emerging challenges, which involve management of renewables and flexibility in energy markets and networks under uncertainty --- forecasting is only part of the solution. Intraday and medium-term (days- to weeks-ahead) horizons have received relatively little attention in past competitions and may benefit from the attention and learning opportunities that competitions provide.
The number and scale of data science competitions are growing in the energy sector as they have become established as a cost-effective means of producing research and development by both public and private organisations. See, for example, Eesti Energia's \citep{Eljand2023EnefitProsumers}, RTE's \textit{Learn to run a power network}~\citep{marot21a} and the ARPA-E \textit{Grid Optimization Competition}~\citep{Aravena2022ARPAE}. Platforms for continuous evaluation and remuneration of energy forecasts are also emerging, such as the Solar Forecast Arbiter\footnote{https://forecastarbiter.epri.com (Accessed 29/11/2024)} and
analytics markets such as Elia's Predico initiative\footnote{https://innovation.eliagroup.eu/en/projects/predico-collaborative-forecasting-platform (Accessed 26/11/2024)}. Furthermore, the recent European Union legislation on artificial intelligence, the \gls{ai} Act, emphasises the need for testing and experimentation facilities (TEFs) as essential infrastructures for assessing the conformity of data-driven models. Operational competitions, such as HEFTcom, conducted in controlled environments, offer valuable insights into evaluation methodologies and metrics, helping to align TEF's practices with the regulatory framework.

HEFTcom has showcased the potential for competitions to tackle practical challenges in renewable energy integration beyond forecasting alone, particularly in decision-making under forecast uncertainty. As the energy sector and artificial intelligence continue to advance alongside evolving regulatory frameworks, future competitions should prioritize emerging needs such as explainability, human-computer interaction, robustness, and the delivery of more prescriptive outputs. Leveraging advancements in open and synthetic data, and open software, will be essential to maximizing their impact.

\section*{Acknowledgements}

We would like to thank \O{}rsted for sponsoring the HEFTcom prize pot, rebase.energy for providing the competition platform and access to numerical weather prediction data, and the IEEE DataPort for hosting the competition.
\ifblind
[Further achnoledgements omitted for blind peer review.]
\else
We would also like to thank Alex Louden, Pierre Pinson and Klimis Stylpnopoulos, and members of the IEA Wind Task 51, for the advice and support they provided.
\fi
This research did not receive any specific grant from funding agencies in the public, commercial, or not-for-profit sectors.

This paper contains data provided by Elexon (BSC information licensed under the BSC Open Data Licence), the Low Carbon Contracts Company (public sector information licensed under the Open Government Licence v3.0), and Sheffield Solar PV Live. Further details and the data itself are available at \citep{Browell2024HybridData}. Code to reproduce the analysis presented in this paper is available on GitHub\footnote{https://github.com/jbrowell/HEFTcom24-Analysis} and \citep{Browell2024HEFTcomAnalysis}.

For the purpose of open access, the authors have applied a Creative Commons Attribution (CC BY) licence to any Author Accepted Manuscript version arising from this submission.

\ifblind
\else
\section*{CRediT author statement}
\textbf{Jethro Browell}: Conceptualization, Methodology, Software, Data Curation, Validation, Investigation, Formal analysis, Writing -- Original Draft, Supervision, Project administration, Funding acquisition;
\textbf{Dennis van der Meer}: Investigation, Supervision, Formal analysis, Writing -- Original Draft;
\textbf{Henrik K\"alvegren}: Software, Data Curation;
\textbf{Sebastian Haglund}: Conceptualization, Supervision, Writing -- Review \& Editing;
\textbf{Edoardo Simioni}: Conceptualization, Supervision;
\textbf{Ricardo J. Bessa}: Supervision, Writing -- Review \& Editing;
\textbf{Yi Wang}: Supervision.
\fi

\section*{Appendix}

Full results from HEFTcom, missed submissions and student status are provided in Table~\ref{tab:full_results}. All competition data is available in \citep{Browell2024HybridData}.

\begin{table}[!ht]
\caption{Full results for HEFTcom ordered by Pinball score. Only teams that submitted reports and missed five or fewer submissions were eligible for a final rank. Teams \textit{Benchmark} and \textit{quantopia} were managed by the organisers and did not receive a rank. \blue{$^\ast$ProbProfit made a single but consequential forecast error without which their average Pinball Score would have been 29.47 MWh.}}
\label{tab:full_results}
\centering
\resizebox{\textwidth}{!}{%
\begin{tabular}{lrrrrrlrl}
  \hline
Team & Pinball [MWh] & Revenue [\pounds m] & Forecasting rank & Trading rank & Combined rank & Report & Missed submissions & Student \\ 
  \hline
SVK & 22.18 & 88.88 & 1 & 1 & 1 & TRUE & 0 &  \\ 
  UI BUD & 23.18 & 88.07 & 2 & 4 & 3 & TRUE & 0 &  \\ 
  Rnt & 24.64 & 88.29 & 3 & 2 & 2 & TRUE & 1 &  \\ 
  GEB & 25.16 & 88.18 & 4 & 3 & 4 & TRUE & 0 & TRUE \\ 
  BridgeForCast & 25.34 & 87.67 & 5 & 6 & 5 & TRUE & 1 &  \\ 
  quantopia & 25.38 & 87.96 &  &  &  & TRUE & 1 &  \\ 
  LSEG Power Team & 25.74 & 85.71 & 6 & 13 & 9 & TRUE & 0 &  \\ 
  sukantabasu & 27.04 & 87.83 & 7 & 5 & 6 & TRUE & 1 &  \\ 
  Stochastic Parrots & 27.50 & 87.53 & 8 & 8 & 7 & TRUE & 1 &  \\ 
  EnergiWise & 27.65 & 87.43 & 9 & 9 & 8 & TRUE & 0 &  \\ 
  NICE\_Forecast & 27.98 & 87.21 & 10 & 11 & 11 & TRUE & 0 & TRUE \\ 
  Oracle & 28.34 & 87.20 & 11 & 12 & 12 & TRUE & 0 &  \\ 
  Ihubex & 29.22 & 87.64 & 12 & 7 & 10 & TRUE & 2 & TRUE \\ 
  RE-Cast & 30.04 & 87.31 & 13 & 10 & 13 & TRUE & 0 &  \\ 
  (Please hug emoji) & 31.01 & 84.29 & 14 & 18 & 15 & TRUE & 1 &  \\ 
  PI9 & 31.12 & 85.89 &  &  &  &  & 2 &  \\ 
  GM Team Mannheim & 33.92 & 85.55 & 15 & 15 & 14 & TRUE & 0 &  \\ 
  Zzblu & 35.04 & 84.41 & 16 & 17 & 16 & TRUE & 3 &  \\ 
  Eguzkinet & 36.68 & 83.83 & 17 & 21 & 19 & TRUE & 5 &  \\ 
  tradRES & 37.19 & 84.45 & 18 & 16 & 17 & TRUE & 4 & TRUE \\ 
  NAECO Blue GmbH & 38.90 & 85.97 &  &  &  & TRUE & 6 &  \\ 
  justForFun & 39.50 & 85.18 &  &  &  & TRUE & 6 &  \\ 
  KittenKilowatt & 41.67 & 83.71 & 19 & 22 & 21 & TRUE & 0 &  \\ 
  OLPZR & 44.84 & 85.58 & 20 & 14 & 18 & TRUE & 0 &  \\ 
  SiaPartners\_Team & 45.23 & 85.05 &  &  &  &  & 2 &  \\ 
  6340 & 47.37 & 84.09 & 21 & 19 & 20 & TRUE & 5 &  \\ 
  CRL & 47.73 & 84.23 &  &  &  & TRUE & 40 &  \\ 
  ReWind & 48.80 & 87.50 &  &  &  & TRUE & 6 &  \\ 
  RUPowered & 48.98 & 83.22 &  &  &  &  & 12 &  \\ 
  Wu Forecast & 49.03 & 83.31 &  &  &  &  & 75 &  \\ 
  ODC & 49.19 & 83.64 &  &  &  & TRUE & 51 & TRUE \\ 
  Energon Unlimited & 50.26 & 83.25 &  &  &  &  & 2 &  \\ 
  The Onliners & 50.39 & 84.01 &  &  &  &  & 63 &  \\ 
  Team Auckland & 50.41 & 82.25 &  &  &  &  & 86 &  \\ 
  Amp-Q & 50.80 & 83.36 &  &  &  &  & 18 &  \\ 
  FCOR\_BL & 51.66 & 82.26 & 22 & 24 & 23 & TRUE & 2 &  \\ 
  TThursday & 51.89 & 82.67 &  &  &  &  & 7 &  \\ 
  Neo & 52.05 & 82.34 &  &  &  &  & 37 &  \\ 
  Matrix & 52.32 & 82.60 &  &  &  &  & 21 &  \\ 
  DDDelft & 52.37 & 81.70 &  &  &  &  & 20 &  \\ 
  HelloWorld & 52.39 & 82.37 &  &  &  &  & 20 &  \\ 
  Intelligent Electrical Power Traders & 52.70 & 79.89 &  &  &  & TRUE & 21 &  \\ 
  ALO-Forecast & 52.93 & 83.85 &  &  &  & TRUE & 10 & TRUE \\ 
  HyForecast & 52.93 & 84.01 &  &  &  &  & 49 &  \\ 
  Auror & 53.09 & 82.54 &  &  &  &  & 76 &  \\ 
  KIT-IAI & 53.43 & 82.41 &  &  &  &  & 84 &  \\ 
  Njord & 53.52 & 83.82 &  &  &  &  & 49 &  \\ 
  cld & 53.55 & 81.97 &  &  &  &  & 53 &  \\ 
  gsp23 & 53.56 & 82.39 &  &  &  &  & 89 &  \\ 
  Benchmark & 53.58 & 82.23 &  &  &  & TRUE & 0 &  \\ 
  aisopb & 53.58 & 82.23 &  &  &  &  & 89 &  \\ 
  power\_rabbit & 53.96 & 84.03 & 23 & 20 & 22 & TRUE & 4 & TRUE \\ 
  vishleshak & 54.24 & 82.04 &  &  &  &  & 81 &  \\ 
  mariscos & 55.05 & 82.23 &  &  &  & TRUE & 56 &  \\ 
  Enerweb & 55.13 & 83.22 & 24 & 23 & 24 & TRUE & 4 &  \\ 
  HJ Energy & 56.18 & 81.15 &  &  &  & TRUE & 81 &  \\ 
  MiaoMiaoJiao & 57.66 & 81.63 &  &  &  &  & 32 &  \\ 
  mizu & 58.54 & 81.68 &  &  &  &  & 0 &  \\ 
  TÃ¡Sol & 60.11 & 79.41 &  &  &  &  & 49 &  \\ 
  CuriousEngineer & 62.49 & 78.53 &  &  &  &  & 29 &  \\ 
  Glassbowl-Prediction & 62.81 & 78.09 &  &  &  & TRUE & 39 &  \\ 
  ForMare & 69.89 & 78.93 &  &  &  & TRUE & 42 &  \\ 
  Aphelion & 71.84 & 71.39 &  &  &  &  & 7 &  \\ 
  CUFE & 77.83 & 69.99 &  &  &  &  & 37 &  \\ 
  forecaaaaast & 108.32 & 56.97 &  &  &  &  & 2 &  \\ 
  ProbProfit & 2645715638.85$^\ast$ & 87.52 &  &  &  &  & 0 &  \\ 
   \hline
\end{tabular}
}
\end{table}

\newpage

\bibliographystyle{elsarticle-harv} 
\bibliography{jethro-mendeley}


\end{document}